\documentclass[conference]{IEEEtran}

\usepackage{amsmath,amssymb}
\usepackage{amsthm}

\usepackage[ruled,vlined]{algorithm2e}

\usepackage{graphicx}
\usepackage{booktabs}
\usepackage{cite}
\usepackage{subfig}

\theoremstyle{definition}

\theoremstyle{plain}

\usepackage{xcolor}

\def\BibTeX{{\rm B\kern-.05em{\sc i\kern-.025em b}\kern-.08em
    T\kern-.1667em\lower.7ex\hbox{E}\kern-.125emX}}

\begin{document}

\title{LLM-Assisted Coalition Formation for Cooperative Perception in Autonomous Driving}

\author{
\IEEEauthorblockN{Ahmad Sarlak, Hao Wang, Rahul Amin, Abolfazl Razi}
\IEEEauthorblockA{
\textit{School of Computing, Clemson University, Clemson, SC, USA}\\
\textit{Lincoln Laboratory, Massachusetts Institute of Technology, Lexington, MA, USA} \\
asarlak@g.clemson.edu, hao9@g.clemson.edu,  rahul.amin@ll.mit.edu, arazi@clemson.edu}
}

\maketitle

\begin{abstract}

Cooperative perception (CP) enables connected autonomous vehicles (CAVs) to share complementary observations for safer navigation, but practical deployment is limited by bandwidth constraints, unreliable links, and redundant information exchange. Existing CP methods often assume predefined participants and merely focus on collective perception. Likewise, recent LLM-based cooperative driving frameworks facilitate multi-vehicle reasoning but do not regulate participation criteria to select more beneficial vehicles. To bridge this gap, we propose an LLM-assisted coalition formation framework that selects the most informative helper vehicles before LLM reasoning. The approach jointly optimizes perceptual diversity using a determinantal point process (DPP) over multimodal vehicle embeddings and communication-aware reliability. This leads to a joint coalition selection and power allocation problem, which we solve efficiently via a relaxed convex reformulation and an ADMM-based optimization strategy that decouples diversity-aware selection from network-aware resource allocation. The selected coalition is then summarized and provided with an LLM reasoning module for efficient and less redundant multi-vehicle decision support. Experimental results show that our approach outperforms other baselines in overall coalition value, while maintaining high diversity and improved networking efficiency. The framework achieves a better balance between task performance and safety across OPV2V and V2V4Real datasets, demonstrating its effectiveness for cooperative autonomous driving with communication constraints.

\end{abstract}

\begin{IEEEkeywords}
cooperative perception, DPP, autonomous driving, LLM-assisted systems 
\end{IEEEkeywords}

\section{Introduction}

Autonomous driving has advanced rapidly with deep neural perception models, multimodal representations, and large-scale driving datasets \cite{sarlak2023diversity}. However, safe driving in complex traffic environments remains challenging when a single connected autonomous vehicle (CAV) faces occlusion, limited sensing range, and varying wireless conditions \cite{sarlak2021approach}. Cooperative perception can alleviate this issue by allowing nearby vehicles to share observations and improve scene awareness. Yet, in practical vehicular networks, communication resources are limited, links are unreliable, and transmitting redundant information from many vehicles may increase latency and energy consumption without significantly improving perception quality \cite{sarlak2025extended}.
Existing work on cooperative autonomous driving has mainly focused on perception and feature fusion for tasks such as detection and tracking \cite{chen2019cooper}. Although these methods improve visibility beyond the sensing range of an individual vehicle, they often assume that participating vehicles are already given or selected using fixed or heuristic strategies \cite{tan2026confidence}. In realistic traffic scenarios, however, neighboring vehicles do not contribute equally useful information: some offer redundant viewpoints, while others provide complementary observations that are crucial for resolving occlusions. At the same time, recent LLM-based cooperative driving frameworks, such as V2V-LLM \cite{chiu2025v2v}, have shown the potential of using LLMs for multi-vehicle reasoning and decision support.
However, existing cooperative perception methods focus on perception and typically assume predefined participants, while LLM-based cooperative driving frameworks reason over already collected multi-vehicle information. As a result, neither line of work addresses communication-aware helper selection before reasoning \cite{sarlak2026reliability}. To address this gap, we propose an LLM-assisted coalition formation framework that selects the most informative helper vehicles before LLM-based reasoning under communication constraints. 

Instead of aggregating information from all neighboring vehicles, the proposed framework selects the most informative helper vehicles based on both data diversity and communication efficiency. Each vehicle encodes its local multimodal observations into compact feature embeddings, and candidate coalitions are evaluated through a utility function that jointly captures perceptual diversity and communication efficiency. To promote complementary and non-redundant observations, we employ a determinantal point process (DPP)-based objective over normalized feature embeddings. To ensure practical cooperation, we further incorporate reliability-aware networking metrics, including throughput, retransmission effects, transmission distance, and power consumption. This results in a constrained joint optimization problem over coalition selection and transmit power. The selected coalition is then structured and provided to the LLM for reasoning and decision-making.

Since the resulting problem is mixed-integer and nonlinear, direct optimization becomes intractable as the number of candidate vehicles increases. We therefore develop a scalable solution based on variable relaxation, log-domain power transformation, and the alternating direction method of multipliers (ADMM), which separates diversity-aware vehicle selection from network-aware resource allocation. Our framework uses the LLM as a reasoning module operating on structured summaries from the selected coalition, including scene coverage, detected objects, and environmental awareness. The proposed method therefore bridges communication-aware coalition optimization and high-level LLM reasoning, enabling more efficient, reliable, and non-redundant cooperative autonomous driving.

The rest of the paper is organized as follows. Section \ref{Related_work} reviews related work. Section \ref{sec:system_model} presents the system model. Section \ref{sec:performance_metrics} introduces the performance metrics and problem formulation. Section \ref{sec:admm} describes the proposed ADMM-based optimization method. Section \ref{sec:simulation} provides simulation results, followed by conclusions.

\section{Related Work}
\label{Related_work}
\subsection{Cooperative Perception in Autonomous Driving}
Cooperative perception has progressed from early feature-sharing and intermediate-fusion methods to more structured and communication-efficient multi-agent perception. Early work such as F-Cooper demonstrated feature-level sharing for connected vehicles \cite{chen2019f}, while V2VNet introduced message passing for joint perception and prediction \cite{wang2020v2vnet}. Later methods such as V2X-ViT and CoBEVT improved multi-agent fusion with attention and transformer-based designs across vehicles and infrastructure \cite{xu2022cobevt, xu2022v2x}. More recent studies have shifted toward communication efficiency:

Where2comm transmits spatially critical regions, while bandwidth-efficient communication modeling learns communication partners and informative content under bandwidth constraints \cite{hu2022where2comm}. Confidence-V2X uses confidence-guided sparse communication and gating, and optimized cooperative perception under imperfect communication studies helper selection and resource allocation under non-ideal links \cite{tan2026confidence}. Overall, this line of work has substantially improved cooperative perception quality and bandwidth–accuracy tradeoffs, but it remains centered primarily on perception performance and does not explicitly address communication-aware helper selection under resource constraints.

\subsection{LLM-based Autonomous Driving}
In parallel, LLM-based autonomous driving has emerged as a promising direction for scene understanding, reasoning, and planning. DriveGPT4 and LMDrive showed that large language models can support interpretable or instruction-following end-to-end driving \cite{xu2024drivegpt4, shao2024lmdrive}, while DriveLM introduced graph-structured visual question answering over perception, prediction, and planning \cite{sima2024drivelm}. More recent multimodal driving models such as EMMA and DriveGPT4-V2 further unify multiple driving tasks within a common reasoning framework \cite{xu2025drivegpt4, hwang2024emma}. In cooperative settings, V2V-LLM demonstrated that a centralized language model can reason over perception features shared by multiple connected vehicles, and V2V-LLM extended this direction with graph-of-thought reasoning for cooperative perception, prediction, and planning \cite{chiu2025v2v}. Collectively, these works highlight the growing strength of LLM-based driving systems, but most operate after multi-vehicle information has already been collected or fused, without considering which vehicles should participate under communication constraints.

\section{System Model}
\label{sec:system_model}

Consider a set of candidate helper vehicles $\mathcal{N}=\{1,\ldots,N\}$.  
Each vehicle $i\in\mathcal{N}$ encodes its local multimodal observations into a
$d$-dimensional feature vector $\mathbf{f}_i\in\mathbb{R}^d$. The corresponding
unit-norm embedding is
\begin{equation}
    \tilde{\mathbf{f}}_i
    = \frac{\mathbf{f}_i}{\|\mathbf{f}_i\|},
    \qquad i\in\mathcal{N}.
\end{equation}

A coalition $S\subseteq\mathcal{N}$ of size $k=|S|$ shares its embeddings with a central aggregator (e.g., ego vehicle or edge server) to support cooperative perception and downstream LLM
reasoning. For $S=\{i_1,\ldots,i_k\}$, define
\begin{equation}
    \tilde{\mathbf{F}}_S
    = \big[\tilde{\mathbf{f}}_{i_1},\tilde{\mathbf{f}}_{i_2},\ldots,\tilde{\mathbf{f}}_{i_k}\big]
    \in\mathbb{R}^{d\times k}.
\end{equation}

Each vehicle $i\in S$ transmits a payload of $L_i$ bits over a wireless link of
achievable rate $R_i$ using power $P_i$. The nominal airtime is
\begin{equation}
    T_i = \frac{L_i}{R_i}.
\end{equation}

Let $\mathbf{P}^{\mathrm{tx}}=[P_1,\ldots,P_N]^\top$ collect the transmit
powers. The utility of coalition $S$ under power allocation
$\mathbf{P}^{\mathrm{tx}}$ is defined as
\begin{equation}
    v\!\left(S,\mathbf{P}^{\mathrm{tx}}\right)
    = w_a\,a(S) + w_b\,b\!\left(S,\mathbf{P}^{\mathrm{tx}}\right),
    \label{coalition_object}
\end{equation}

where $w_a, w_b \ge 0$ and $w_a + w_b = 1$ denote the relative importance of the two objectives, namely $a(S)$ measures perceptual diversity, 
and $b(S,\mathbf{P}^{\mathrm{tx}})$ networking efficiency, defined in equations Eq.(\ref{eq:a_s}) and Eq.(\ref{eq:bS_def}) respectively. Transmit power variables are defined for all candidate vehicles, while only vehicles in the selected coalition $S$ actively transmit. The transmit power is subject to the following constraints:

\begin{align}
    0 < P_i &\le P_{\max}, \qquad \forall i\in\mathcal{N},
    \label{eq:power_constraint} \\
    \sum_{i\in S} P_i &\le P_{\mathrm{tot}},
    \label{eq:total_power_constraint}
\end{align}
where $P_{\max}$ is the per-vehicle transmit-power limit and
$P_{\mathrm{tot}}$ is the total power budget.

\begin{figure}[t]
\centering
\includegraphics[width=\columnwidth]{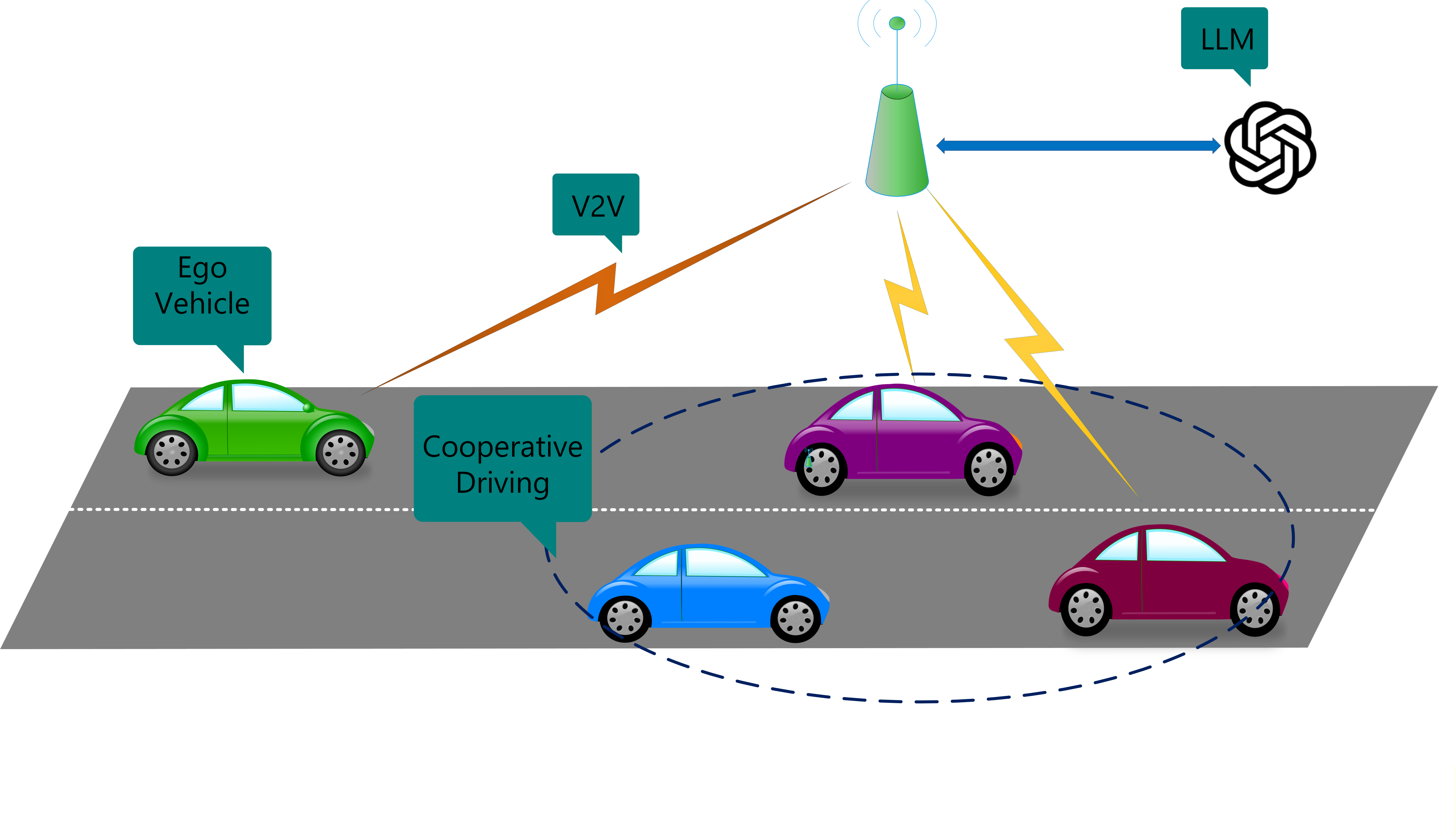}
\caption{LLM-assisted coalition formation for cooperative driving.}
\label{fig:R2}
\end{figure}

The objective is to select a coalition that maximizes information diversity while efficiently utilizing communication resources.

\section{Performance Metrics and Problem Formulation}
\label{sec:performance_metrics}

We next define the metrics used to quantify coalition quality.
\subsection{Spectral Diversity}

We represent coalition membership by a binary selection vector
$\boldsymbol{\gamma}\in\{0,1\}^N$ satisfying
\begin{equation}
    \mathbf{1}^\top \boldsymbol{\gamma} = k,
\end{equation}
where $\mathbf{1}$ is the all-ones vector. The corresponding coalition is
\begin{equation}
    S(\boldsymbol{\gamma})
    = \{i\in\mathcal{N}:\gamma_i=1\}.
\end{equation}

Collect all normalized embeddings in
\begin{equation}
    \tilde{\mathbf{F}}
    = \big[\tilde{\mathbf{f}}_1,\tilde{\mathbf{f}}_2,\ldots,\tilde{\mathbf{f}}_N\big]
    \in\mathbb{R}^{d\times N}.
\end{equation}
For a given coalition $S$, let $\tilde{\mathbf{F}}_S$ denote the submatrix of
$\tilde{\mathbf{F}}$ indexed by $S$ where each column corresponds to a selected vehicle in $S$. and define the Gram matrix
\begin{equation}
    \mathbf{K}_{S,S}
    = \tilde{\mathbf{F}}_S^\top \tilde{\mathbf{F}}_S
    \in\mathbb{R}^{k\times k}.
\end{equation}

We measure coalition diversity using the DPP-inspired score
\begin{equation}
    a(S) \triangleq \log\det\!\left(\mathbf{K}_{S,S}\right).
    \label{eq:a_s}
\end{equation}
This objective favors coalitions whose embeddings span a large volume in the embedding space, and thus
capture complementary rather than redundant observations. The corresponding
size-$k$ diversity maximization problem is
\begin{align}
    \max_{\boldsymbol{\gamma}\in\{0,1\}^N}
    \quad & a\!\left(S(\boldsymbol{\gamma})\right)
    \label{eq:dpp_obj} \\
    \text{s.t.} \quad
    & \mathbf{1}^\top \boldsymbol{\gamma} = k.
    \label{eq:dpp_constraint}
\end{align}

\subsection{Networking Efficiency}

To account for communication quality, we model the reliability of the link from
vehicle $i$ to the aggregator as
\begin{equation}
    \bar{\alpha}_i(P_i)
    = \exp\!\left(-c\,\frac{d_i^{\eta}}{P_i}\right),
\end{equation}
where $d_i$ is the transmission distance, $\eta$ is the path-loss exponent, and
$c>0$ absorbs the effects of noise and interference. The corresponding 
error probability is
\begin{equation}
    \epsilon_i(P_i)
    = 1-\bar{\alpha}_i(P_i).
\end{equation}

The effective throughput of vehicle $i$ is
\begin{equation}
    \zeta_i(P_i)
    = R_i\big(1-\epsilon_i(P_i)\big),
    \label{eq:throughput_i}
\end{equation} 

Under geometric retransmissions (i.e., repeated transmissions until success), the expected transmission energy is
\begin{equation}
    E_i(P_i)
    = \frac{P_i L_i}{1-\epsilon_i(P_i)},
    \label{eq:energy_i}
\end{equation}

We combine the two into a coalition-level networking score:
\begin{equation}
    b\!\left(S,\mathbf{P}^{\mathrm{tx}}\right)
    = \frac{1}{|S|} \sum_{i\in S}
    \left( w_T\, \zeta_i(P_i) - w_E\, E_i(P_i) \right),
    \label{eq:bS_def}
\end{equation}

where $w_T,w_E\ge 0$ and $w_T+w_E=1$. A coalition achieves a high networking score if it
achieves high throughput with low energy consumption.

\subsection{Problem Formulation}

Given the diversity metric $a(S)$ and the networking metric
$b(S,\mathbf{P}^{\mathrm{tx}})$, the joint coalition selection and power
allocation problem is
\begin{align}
    \max_{\boldsymbol{\gamma},\,\mathbf{P}^{\mathrm{tx}}}
    \quad &
    v\!\left(S(\boldsymbol{\gamma}),\mathbf{P}^{\mathrm{tx}}\right)
    \label{eq:opt_obj} \\
    \text{s.t.} \quad
    & \mathbf{1}^\top \boldsymbol{\gamma} = k,
    \label{eq:opt_cardinality} \\
    & 0 < P_i \le P_{\max}, \qquad \forall i\in\mathcal{N},
    \label{eq:opt_power} \\
    & \sum_{i=1}^{N} \gamma_i P_i \le P_{\mathrm{tot}},
    \label{eq:opt_total_power} \\
    & \gamma_i \in \{0,1\}, \qquad \forall i\in\mathcal{N}.
    \label{eq:opt_discrete}
\end{align}

Problem~\eqref{eq:opt_obj}--\eqref{eq:opt_discrete} is a mixed-integer
nonlinear program (MINLP). The binary vector $\boldsymbol{\gamma}$ controls
coalition composition and perceptual diversity, while
$\mathbf{P}^{\mathrm{tx}}$ determines the communication efficiency of the
selected coalition. Direct solution is intractable for large candidate sets;
therefore, we next develop a scalable ADMM-based solver.

\section{ADMM-Based Coalition Optimization}
\label{sec:admm}

To enable scalable optimization, we reformulate the problem as a continuous optimization problem and solve it using ADMM.

To obtain a tractable approximation of
\eqref{eq:opt_obj}--\eqref{eq:opt_discrete}, we relax the binary variables to
\begin{equation}
    0 \le \gamma_i \le 1,
    \qquad
    \mathbf{1}^\top \boldsymbol{\gamma} = k.
\end{equation}
Under this relaxation, the diversity term $a(S(\boldsymbol{\gamma}))$ is interpreted via a continuous log-determinant surrogate in $\boldsymbol{\gamma}$. We also introduce the log-power variables
\begin{equation}
    y_i = \log P_i,
    \qquad
    P_i = e^{y_i},
\end{equation}
which convert the power constraints into
\begin{align}
    y_i &\le \log P_{\max}, \qquad \forall i,
    \label{eq:log_power_box} \\
    &\sum_{i=1}^{N} \gamma_i e^{y_i} \le P_{\mathrm{tot}}.
    \label{eq:log_power_sum}
\end{align}

Under this transformation, the DPP term admits a standard continuous
log-determinant relaxation in $\boldsymbol{\gamma}$, while the negative
networking objective admits a convex approximation in $\mathbf{y}$ under the adopted link model. Let
\begin{equation}
    b(\boldsymbol{\gamma},\mathbf{y})
    \triangleq
    \frac{w_T}{k}\sum_{i=1}^{N}\gamma_i\,\zeta_i(e^{y_i})
    - \frac{w_E}{k}\sum_{i=1}^{N}\gamma_i\,E_i(e^{y_i}).
\end{equation}
The relaxed problem can then be written as
\begin{equation}
    \min_{\boldsymbol{\gamma},\,\mathbf{y}}
    \;
    -w_a\,a\!\left(S(\boldsymbol{\gamma})\right)
    -w_b\,b(\boldsymbol{\gamma},\mathbf{y})
\end{equation}
subject to \eqref{eq:log_power_box}, \eqref{eq:log_power_sum}, and
$0\le\gamma_i\le 1$, $\mathbf{1}^\top\boldsymbol{\gamma}=k$.

Define the sets
\begin{align}
    \Gamma
    &= \Big\{
        \boldsymbol{\gamma}\in\mathbb{R}^N:
        0\le\gamma_i\le 1,\;
        \mathbf{1}^\top\boldsymbol{\gamma}=k
    \Big\}, \\
    \mathcal{Y}
    &= \Big\{
        \mathbf{y}\in\mathbb{R}^N:
        y_i \le \log P_{\max},\;
        \sum_{i=1}^{N} \gamma_i e^{y_i} \le P_{\mathrm{tot}}
    \Big\}.
\end{align}
Let $I_{\mathcal{A}}(\cdot)$ denote the indicator function of a set
$\mathcal{A}$. We split the objective into
\begin{align}
    f(\boldsymbol{\gamma})
    &= -w_a\,a\!\left(S(\boldsymbol{\gamma})\right)
    + I_{\Gamma}(\boldsymbol{\gamma}), \\
    g(\boldsymbol{\gamma},\mathbf{y})
    &= -w_b\,b(\boldsymbol{\gamma},\mathbf{y})
    + I_{\mathcal{Y}_{(\gamma)}}(\mathbf{y}).
\end{align}

To enable variable splitting, we introduce two copies of the decision vector,
$x=(\boldsymbol{\gamma}_x,\mathbf{y}_x)$ and
$z=(\boldsymbol{\gamma}_z,\mathbf{y}_z)$, we obtain
\begin{align}
    \min_{x,z}
    \quad & f(\boldsymbol{\gamma}_x)
    + g(\boldsymbol{\gamma}_z,\mathbf{y}_z) \\
    \text{s.t.} \quad & x-z=0.
\end{align}

Using the scaled dual variable $u$, the augmented Lagrangian is
\begin{equation}
    \mathcal{L}_{\rho}(x,z,u)
    = f(\boldsymbol{\gamma}_x)
    + g(\boldsymbol{\gamma}_z,\mathbf{y}_z)
    + \frac{\rho}{2}\|x-z+u\|_2^2
    - \frac{\rho}{2}\|u\|_2^2.
\end{equation}

The ADMM iterations are
\begin{align}
    x^{t+1}
    &= \arg\min_x
    \left[
        f(\boldsymbol{\gamma}_x)
        + \frac{\rho}{2}\|x-z^t+u^t\|_2^2
    \right], \label{eq:admm_x} \\
    z^{t+1}
    &= \arg\min_z
    \left[
        g(\boldsymbol{\gamma}_z,\mathbf{y}_z)
        + \frac{\rho}{2}\|x^{t+1}-z+u^t\|_2^2
    \right], \label{eq:admm_z} \\
    u^{t+1}
    &= u^t + x^{t+1}-z^{t+1}.
    \label{eq:admm_u}
\end{align}

The $x$- and $z$-updates are solved by projected gradient descent (PGD). Let
$F_x(x)$ and $F_z(z)$ denote the objectives in
\eqref{eq:admm_x} and \eqref{eq:admm_z}, respectively. For each variable $q \in \{x,z\}$, the PGD update is
\begin{align}
    q^{\ell+\frac{1}{2}}
    &= q^\ell - \alpha \nabla F_q(q^\ell), \\
    q^{\ell+1}
    &= \Pi_{\mathcal{C}_q}\!\left(q^{\ell+\frac{1}{2}}\right),
\end{align}
where $\Pi_{\mathcal{C}_q}(\cdot)$ denotes Euclidean projection onto the
feasible set. For the $x$-update,
$\mathcal{C}_x=\Gamma\times\mathbb{R}^N$, so the projection reduces to
projecting $\boldsymbol{\gamma}$ onto the capped simplex $\Gamma$ while leaving
$\mathbf{y}$ unchanged. For the $z$-update,
$\mathcal{C}_z=\mathbb{R}^N\times\mathcal{Y}_{(\gamma)}$, and the projection enforces the
log-power box and sum-power constraints.

The inner PGD loop terminates after a fixed number of iterations
$T_{\mathrm{inner}}$ or when
\begin{equation}
    \frac{\|q^{\ell+1}-q^\ell\|_2}{\|q^\ell\|_2}
    \le \varepsilon_{\mathrm{pgd}}.
\end{equation}
After ADMM converges, the transmit powers are recovered as
$P_i=e^{y_i}$. The final discrete coalition is obtained by selecting the $k$ largest entries of $\boldsymbol{\gamma}$.

\section{Simulation}
\label{sec:simulation}

We evaluate the proposed framework in two settings: 1) synthetic large-scale coalition selection and 2) cooperative driving based on end-to-end LLM on the OPV2V~\cite{xu2022opv2v} and V2V4Real \cite{xu2023v2v4real} datasets. The synthetic studies examine optimization behavior and the trade-off between diversity and networking efficiency under controlled conditions, while the real-world study evaluates the impact of coalition optimization on downstream LLM reasoning.
In the synthetic experiments, we consider a vehicular network with 20 candidate vehicles. Each vehicle is represented by a normalized feature embedding of its local observation, and each communication link is characterized by distance-dependent reliability, channel rate, effective throughput, and energy cost. The proposed method jointly selects a coalition and allocates transmit power subject to coalition-size and transmit power constraints. Unless otherwise stated, all synthetic results are averaged over 100 independent random scenarios.

Table~\ref{tab:method_comparison_diversity} compares the method with k-DPP, KMeans, Random, and Channel-Aware (which selects vehicles based on channel quality) under equal power allocation (i.e., identical transmit power for all selected vehicles), ensuring a fair comparison focused solely on coalition selection. The proposed method achieves the highest overall coalition value $v(S)$ and the best networking score $b(S)$, while maintaining competitive diversity performance. Although k-DPP attains the highest diversity score $a(S)$, its weaker networking efficiency reduces its overall utility. In contrast, our approach provides a more balanced trade-off, achieving diversity close to k-DPP while significantly improving networking performance, leading to the best overall outcome. Other baselines, including KMeans, Random, and Channel-Aware Top-$k$, exhibit lower diversity and/or networking efficiency, resulting in inferior coalition values. These results demonstrate the importance of jointly considering diversity and communication efficiency, even under equal power allocation.

\begin{table}[t]
\vspace{0.06in}
\centering
\small
\begin{tabular}{lccc}
\hline
Method & $v(S)$ $\uparrow$ & $a(S)$ $\uparrow$ & $b(S)$ $\uparrow$ \\
\hline
Proposed & $\mathbf{1.65 \pm 0.22}$ & $0.84 \pm 0.24$ & $\mathbf{0.32 \pm 0.04}$ \\
k-DPP & $1.42 \pm 0.24$ & $\mathbf{1.14 \pm 0.07}$ & $0.15 \pm 0.07$ \\
KMeans & $1.04 \pm 0.27$ & $0.67 \pm 0.21$ & $0.15 \pm 0.06$ \\
Random & $0.90 \pm 0.29$ & $0.55 \pm 0.26$ & $0.14 \pm 0.06$ \\
Channel-Aware & $1.03 \pm 0.32$ & $0.60 \pm 0.23$ & $0.17 \pm 0.07$ \\
\hline
\end{tabular}
\caption{Coalition selection under equal power allocation.}
\label{tab:method_comparison_diversity}
\end{table}


Table~\ref{tab:power_comparison} compares the method with several standard power allocation baselines, including equal power, random feasible allocation, channel inversion, and power-only projected gradient descent (PGD), under fixed coalition structures. The proposed method achieves the highest overall coalition value $v(S)$ and networking score $b(S)$, indicating more effective utilization of communication resources. The performance gain is primarily driven by a substantial reduction in energy consumption, while maintaining competitive throughput levels. In contrast, baseline methods achieve similar throughput but incur significantly higher energy consumption, which limits their overall utility. Among the baselines, power-only PGD improves upon heuristic approaches by reducing energy usage and increasing the networking score; however, it remains notably inferior to the our approach. The proposed approach achieves the lowest energy consumption by a wide margin, with only a modest reduction in throughput, resulting in the best trade-off between communication efficiency and reliability. These results demonstrate the effectiveness of the ADMM-based optimization in balancing throughput and energy under practical communication constraints.

\begin{figure}[htbp]
\centering
\includegraphics[width=0.8\columnwidth]{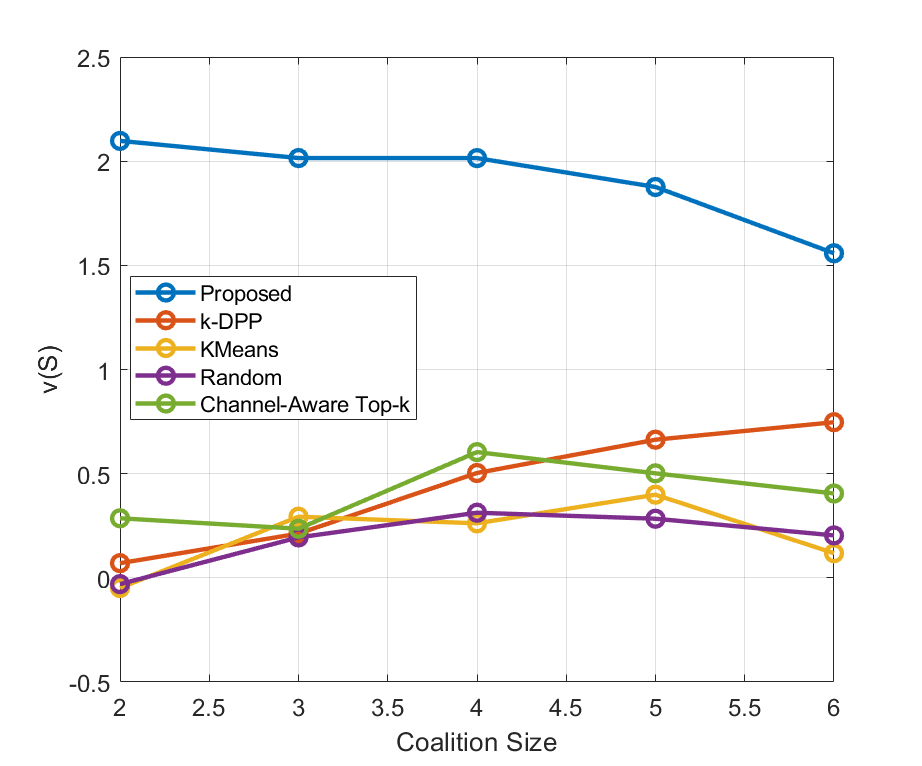}
\caption{Overall coalition value versus coalition size.}
\label{fig:Divers_p_n}
\end{figure}

Fig.~\ref{fig:Divers_p_n} shows the overall coalition value $v(S)$ versus coalition size $k$ under fixed transmit power. The proposed method consistently achieves the highest utility across all coalition sizes. The performance gap widens with $k$, indicating more efficient utilization of cooperative resources. This demonstrates the advantage of the ADMM-based optimization in balancing diversity, communication efficiency, and resource allocation.

\begin{figure}[htbp]
\centering
\subfloat[]{
\includegraphics[width=0.8\columnwidth]{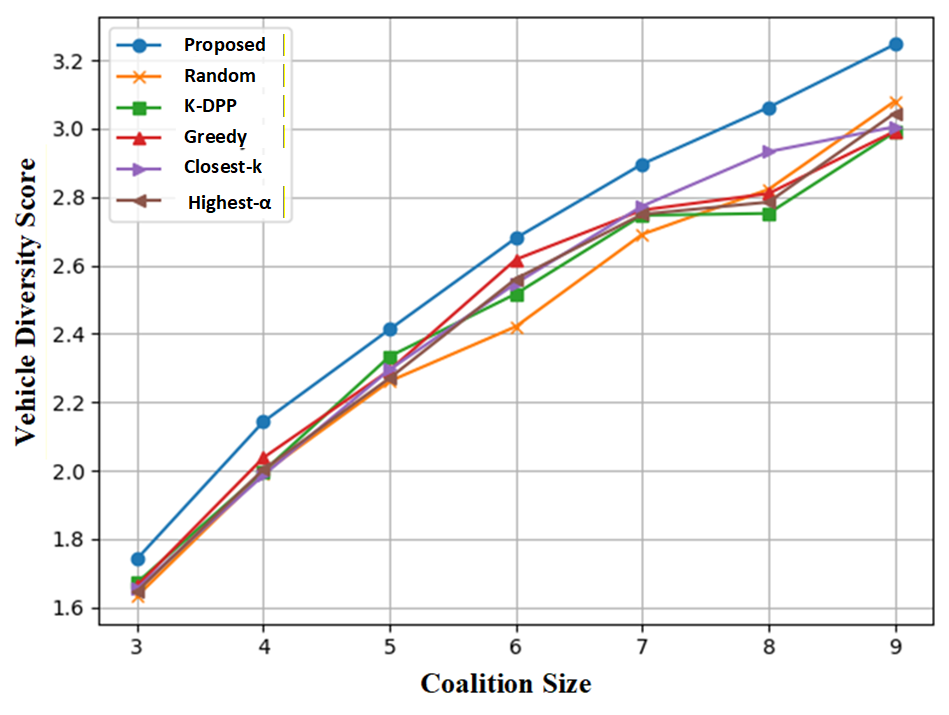}
}
\caption{Diversity under realistic communication conditions.}
\label{fig:real_diversity_networking}
\end{figure}

We next consider a realistic vehicular communication setting with heterogeneous feature embeddings and nonidentical links. In this scenario, coalition selection must jointly balance perceptual diversity and networking efficiency. For comparison, baseline methods use simple heuristics: k-DPP selects vehicles based on diversity with random power allocation, highest-$\alpha$ selects vehicles with the most reliable links, and closest-$k$ selects the nearest vehicles based on distance. As shown in Fig.~\ref{fig:real_diversity_networking}, the proposed method consistently achieves the highest diversity score across coalition sizes.

Finally, we evaluate the proposed framework in an end-to-end LLM-based cooperative driving setting on both the OPV2V and V2V4Real datasets. For each frame, a candidate set of five helper vehicles is constructed, and two helpers are selected using different coalition selection strategies. The selected helper summaries are provided to Llama3, served locally via Ollama with deterministic decoding ($T=0$), to perform structured multi-vehicle reasoning. The LLM input consists of structured summaries derived from multi-vehicle observations, including regional object counts and distance-based spatial features. The model predicts multiple perception labels along with a final driving decision, which is further refined using a rule-based risk score. Ground-truth labels are generated from oracle summaries constructed using the full five-vehicle fused view, ensuring a consistent and full-information reference for evaluation.

We evaluate both decision quality and safety-related behavior using a comprehensive set of metrics. Specifically, We report different metrics. Mean Field Accuracy evaluates per-label prediction accuracy, while Action Accuracy reflects the accuracy of the final driving decision. Unsafe Under-Reaction quantifies the frequency of insufficiently cautious responses in critical situations, whereas Over-Conservative measures unnecessary cautious actions. Additionally, Field Macro-F1 and Action Macro-F1 capture balanced performance across perception fields and action classes, respectively.

Tables~\ref{tab:v2v4real_results} and~\ref{tab:llm_new_results} summarize the results on V2V4Real and OPV2V, respectively. Across both datasets, the proposed method consistently achieves the best overall performance, including higher decision accuracy and improved structured prediction quality. It also maintains competitive or improved safety-related metrics, reducing unsafe under-reaction while avoiding overly conservative behavior. These results demonstrate that communication-aware coalition selection improves both structured scene understanding and downstream decision-making. Overall, our approach provides a more effective balance between performance and safety compared to diversity-based and random selection strategies within the LLM-driven cooperative driving pipeline.

\begin{table*}[!t]
\vspace{0.5in}
\centering
\small
\begin{tabular}{lcccc}
\hline
Method & $v(S)$ $\uparrow$ & $b(S)$ $\uparrow$ & Throughput $\uparrow$ & Energy $\downarrow$ \\
\hline
Equal power & $2.39 \pm 1.63$ & $0.15 \pm 0.18$ & $2.158 \pm 0.34$ & $3.12 \pm 0.63$ \\
Random feasible & $2.11 \pm 1.82$ & $0.12 \pm 0.20$ & $2.10 \pm 0.36$ & $3.18 \pm 0.70$ \\
Channel inversion & $2.44 \pm 1.68$ & $0.16 \pm 0.19$ & $\mathbf{2.15 \pm 0.34}$ & $3.10 \pm 0.71$ \\
Power-only PGD & $2.76 \pm 1.54$ & $0.20 \pm 0.17$ & $2.15 \pm 0.34$ & $2.90 \pm 0.56$ \\
Proposed & $\mathbf{6.22 \pm 1.20}$ & $\mathbf{0.59 \pm 0.13}$ & $2.00 \pm 0.35$ & $\mathbf{0.59 \pm 0.14}$ \\
\hline
\end{tabular}
\caption{Power allocation under fixed coalitions.}
\label{tab:power_comparison}
\end{table*}

\begin{table*}[t]
\centering
\begin{tabular}{l|c c c c c}
\hline
Method & Mean Field Acc (\%) $\uparrow$ & Action Acc (\%) $\uparrow$ & Unsafe Under-Reaction (\%) $\downarrow$ & Over-Conservative (\%) $\downarrow$ \\
\hline
Proposed & \textbf{56} & \textbf{48} & \textbf{52} & 10 \\
Random & 54 & 46 & 52 & \textbf{9} \\
k-DPP & 54 & 45 & 53 & \textbf{9} \\
\hline
\end{tabular}
\caption{Performance of LLM-based cooperative driving on the V2V4Real dataset under different coalition selection strategies.}
\label{tab:v2v4real_results}
\end{table*}

\begin{table}[t]
\centering
\begin{tabular}{l|c c c}
\hline
Metric & Proposed & k-DPP & Random \\
\hline
Action Acc (\%) $\uparrow$ & \textbf{10.77} & 8.90 & 6.10 \\
Unsafe Under-Reaction (\%) $\downarrow$ & \textbf{89.28} & 91.00 & 94.96 \\
Field Macro-F1 (\%) $\uparrow$ & \textbf{25.12} & 24.00 & 12.00 \\
Action Macro-F1 (\%) $\uparrow$ & \textbf{8.36} & 6.23 & 3.28 \\
\hline
\end{tabular}
\caption{Performance comparison of coalition selection strategies on the OPV2V dataset under different helper selection strategies.}
\label{tab:llm_new_results}
\end{table}

\section{Conclusion}

This paper presented an LLM-assisted coalition formation framework for cooperative autonomous driving under communication constraints. Our approach addresses the gap between communication-aware vehicle selection and LLM-based multi-vehicle reasoning by selecting the most informative helper vehicles before reasoning. We formulated joint coalition selection and power allocation as an optimization problem that integrates DPP-based diversity with communication efficiency metrics, and solved it using a scalable ADMM-based approach. Results showed that the proposed method consistently outperforms baseline coalition selection strategies in overall utility, while achieving higher diversity and better networking efficiency. Experiments on OPV2V and V2V4Real datasets further demonstrated improved downstream cooperative driving performance, with a better balance between task accuracy, safety-related behavior, and communication efficiency. These findings suggest that optimized coalition formation is a practical front-end for LLM-based cooperative driving, enabling more reliable and less redundant information sharing in realistic vehicular networks.

\bibliographystyle{IEEEtran}   
\bibliography{cas-refs}      
\end{document}